\documentclass[preprint2]{emulateapj-rtx4}

\shorttitle{Radio observations of GRB 100418a}
\shortauthors{Moin et al.}

\begin{document}

\title{Radio observations of GRB 100418a: Test of an energy injection model explaining long-lasting GRB afterglows}

\author{A. Moin$^{1,2}$, P. Chandra$^{3}$, J.~C.~A. Miller-Jones$^{2}$, S.~J. Tingay$^{2}$, G. B. Taylor$^{4,5}$, D. A. Frail$^{5}$, Z. Wang$^{1}$, C. Reynolds$^{2}$, C.~J. Phillips$^{6}$ \\ $^{1}$Shanghai Astronomical Observatory, 80 Nandan Road, Xujiahui, Shanghai 200030, China \\ $^{2}$International Centre for Radio Astronomy Research, Curtin University, Bentley 6102, WA, Australia \\ $^{3}$National Center for Radio Astrophysics, Tata Institute of Fundamental Research, Pune, India \\ $^{4}$Department of Physics and Astronomy, University of New Mexico, Albuquerque NM, 87131, USA \\ $^{5}$National Radio Astronomy Observatory, , P.O. Box O, Socorro, NM 87801, USA \\ $^{6}$CSIRO Astronomy and Space Science, PO Box 76, Epping, NSW, Australia}

\begin{abstract}

We present the results of our radio observational campaign on GRB 100418a, for which we used the Australia Telescope Compact Array (ATCA), Very Large Array (VLA) and the Very Long Baseline Array (VLBA). GRB 100418a was a peculiar GRB with unusual X-ray and optical afterglow profiles featuring a plateau phase with a very shallow rise. This observed plateau phase was believed to be due to a continued energy injection mechanism which powered the forward shock, giving rise to an unusual and long-lasting afterglow. The radio afterglow of GRB 100418a was detectable several weeks after the prompt emission.  We conducted long-term monitoring observations of the afterglow and attempted to test the energy injection model advocating that the continuous energy injection is due to shells of material moving at a wide range of Lorentz factors. We obtained an upper limit of  $\gamma$ $<$ 7  for the expansion rate of the GRB 100418a radio afterglow, indicating that the range-of-Lorentz factor model could only be applicable for relatively slow moving ejecta. A preferred explanation could be that continued activity of the central engine may have powered the long-lasting afterglow.

\end{abstract}


\section{Introduction}

Gamma-Ray Burst (GRB) 100418a was detected by the Swift satellite at 21:10:08 UT, 2010 April 18. The Burst Alert Telescope (BAT) on board the Swift satellite was triggered by the GRB and after the initial trigger, Swift's X-Ray Telescope (XRT) and Ultra-Violet Optical Telescope (UVOT) slewed to the source and detected it in the X-ray and optical bands \citep{mar11}. Subsequently, a number of ground-based telescopes carried out detailed studies of the afterglow of GRB 100418a. Optical observations with the Very Large Telescope and the Gemini North Telescope determined the redshift of GRB 100418a to be 0.6235 \citep{Ant10, Cuc10}.

\citet{mar11} analyzed the Swift BAT, XRT and UVOT data and reported that the afterglow of the GRB exhibited unusual behaviour in the optical and X-ray bands. After a brief period of steep decline, the X-ray light curve featured an unusually long "plateau" phase, lasting from about 600 seconds until 50--90 ks after the burst.  The optical afterglow also showed a period of rebrightening with a similar power-law index, from about 87 seconds until 51 ks after the burst.  Following this shallow rise in the light curves, the afterglow resumed a steeper and more normal power law decay.

\citet{Jia12} and \citet{Nii12} independently found that there was no evidence of supernova association with GRB 100418a. Together with the high metallicity of the host galaxy \citep{Uga11, Nii12}, the absence of a supernova and a peculiar, long-lasting afterglow leads to the inference that either the progenitor belongs to an entirely different class or that it behaved in a way different from how the GRB afterglow would be expected to behave within the framework of the standard collapsar--fireball model \citep{Mac99, Mac01}.

Long-term multi-wavelength observational studies of the afterglow of peculiar GRBs such as GRB 100418a provide an opportunity to examine the unusual behaviour of afterglows of these rare GRBs, test previously advanced theories of GRB physics \citep{ree92, ree98, Pan98, Wak98, Dai98, Zha06}, and hold the potential to find clues to the nature of GRB central engines. Motivated by the unusual afterglow behaviour, we conducted radio monitoring of GRB 100418a, in an attempt to understand the astrophysical mechanism behind the afterglow profile and the role of the central engine, and to test the "energy injection" model \citep{ree98, Zha06}, which offers a plausible explanation of the physics of long-term afterglows of GRBs. While a long-lasting afterglow could in principle be explained by the jets running into the dense circumburst medium (with the time delay between approaching and receding jets prolonging the afterglow), the lack of an associated supernova led us to focus this work on the energy injection model. In what follows, we show that the unique behaviour of the afterglow of this GRB supplies some observational evidence to support one of the postulates of the energy-injection model, which advocates continued activity of the central engine as the primary driver for long-lasting afterglows.

\section{Observations and Results}

\begin{figure*}
\includegraphics[angle=0,scale=.50]{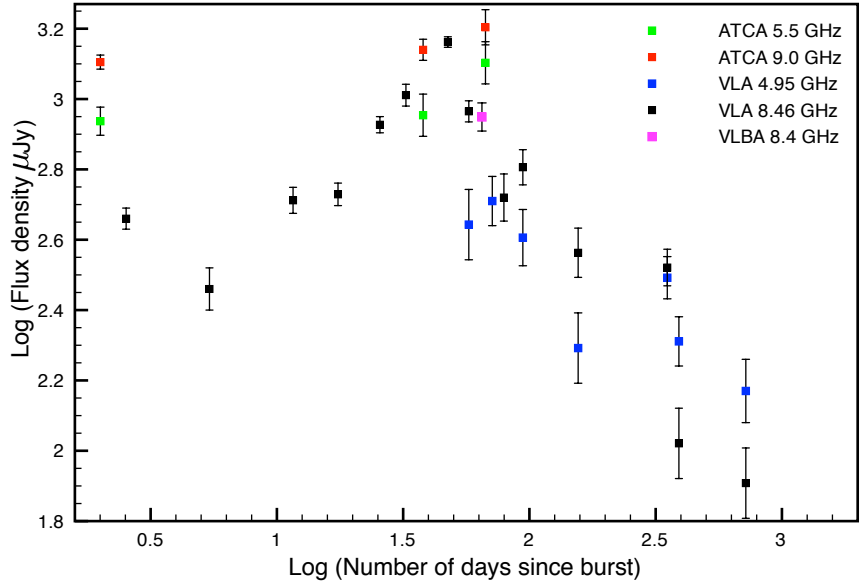}
\centering
\caption{GRB 100418a was detected and monitored from 2010 April until 2012 August by VLA and ATCA in the 4.95 and 8.46 GHz, 5.5 and 9.0 GHz bands respectively. It was also observed with VLBA at 8.4 GHz. This light-curve presents the record of long-term post-burst VLA + ATCA + VLBA radio observations (error bars: 1$\sigma$) of the GRB 100418a afterglow.\label{fig1}}
\end{figure*}

Radio monitoring of GRB 100418a was conducted using the Australia Telescope Compact Array (ATCA) in Australia \citep{Moi10} and the Very Large Array in the United States \citep{Cha10}. Owing to the significant and consistent brightness at radio wavelengths, an observation was conducted with the Very Long Baseline Array (VLBA) to observe the GRB afterglow at high angular resolution. Fig.~\ref{fig1} shows the combined VLA + ATCA light curve for the radio afterglow of GRB 100418a. It also includes the VLBA observation.

\subsection{VLA Observations}

VLA observations of GRB 100418a started 2010 April 21. Until 2010 September 09, the observations were conducted at 4.95 and 8.46 GHz. Subsequently, observations were carried out at 4.9 and 7.9 GHz. The VLA monitoring continued through 2012 August. The data were taken in the interferometric mode for an average of 30 minutes (including calibrator time). 3C286 was used as the flux calibrator as well as bandpass calibrator. J1728+1215 was used for the phase calibration. The data were analyzed using standard AIPS\footnote[1]{http://www.aips.nrao.edu/index.shtml} \citep{Gre03} routines. The flux density of the radio afterglow continued to rise until it reached a peak about 60 days after the prompt emission and the post-peak phase featured a gradual decline as the radio afterglow was fading away. Details of all the radio observations are listed in Table~\ref{obs}.

\subsection{ATCA Observations}

GRB 100418a was observed using the ATCA during three epochs of an ongoing GRB observation and monitoring program (ATCA proposal C1802), conducted in 2010 April, May and June (See Table~\ref{obs}). It was detected at all three epochs at 5.5 and 9.0 GHz. 

The source was observed with a nearby calibrator (J1655+077) for phase referencing and PKS 1934-638 was observed during the session for primary flux and bandpass calibration. Data were taken with the Compact Array Broadband Backend (CABB: \citealt{Wil11}) in the standard continuum mode with a bandwidth of 2.048 GHz, in two polarisations. 

The output data files were then loaded into MIRIAD\footnote[2]{http://www.atnf.csiro.au/computing/software/miriad/} \citep{Sau95} for calibration. Standard MIRIAD calibration routines were used to calibrate the amplitude and the phase gains of the data and FITS files of the calibrated data were produced, which were then loaded into Difmap\footnote[3]{ftp://ftp.astro.caltech.edu/pub/difmap/difmap.html} \citep{She94} to produce final images of the GRB 100418a afterglow.  In each image, the detected afterglow was fitted with a point-source model using the model-fitting routine with five iterations and the peak flux density for each session was determined. The errors on flux densities are the 1-sigma value of the RMS noise.

Even a number of weeks after the burst, it was possible to detect the radio afterglow, as can be seen from Fig. \ref{fig1}, which was quite unusual as the likelihood of detecting the radio afterglow of most GRBs starts to diminish within a few days of the prompt emission. The recent GRB radio afterglow catalog \citep{Cha12} reported that the GRBs with radio afterglows lasting more than seven days comprised approximately 22\% of the entire sample of GRBs observed as part of the GRB radio follow-up campaign running since 1997, and the GRB radio afterglows which were monitored and studied over several months made only approximately 5\% of the full sample.  

The ATCA + VLA light curve features a brief decline before it started to rise from day 11 onwards  until it reached its peak about 60 days after the burst. There were no further ATCA observations of the GRB 100418a radio afterglow after the third session in 2010 June but the VLA continued to monitor it. The spectral indices from the late time observations of the GRB do indicate that the afterglow emission transitioned from inverted to steep spectrum, which is an expected feature of long-lasting radio afterglows at late times. Noticeable short time-scale variations between the ATCA flux densities and those obtained from the VLA epochs are possibly due to the diffractive scintillation effects in the ionized interstellar medium \citep{Fra00}, with additional contribution from the differences in the flux and phase calibrators that were used at both observatories to calibrate the GRB 100418a data.

\subsection{VLBA Observations}

The VLBI observation of GRB 100418a was an attempt to resolve the afterglow and to determine the possible milliarcsecond-scale structure associated with the afterglow. Previous VLBI monitoring of the bright and nearby GRB 030329 provided important clues to the evolution and behavior of GRB radio afterglows \citep{Tay04, Tay05, Pih07}.  In that light, the VLBI observation of GRB 100418a was conducted in an attempt to resolve the afterglow and any associated milliarcsecond-scale structure.  In particular, an estimate of its angular size and expansion rate could help constrain the energy injection model.

VLBA observations of GRB 100418a were carried out as part of program BM347 on 2010 June 22. The source was observed for eight hours with full Stokes parameters (dual polarisation) at a frequency of 8.4 GHz with a bandwidth of 8 $\times$ 8 MHz. Along with the target source, phase-reference calibrator J1706+1208 and the VLBA fringe finder 3C345 were also observed. The data were correlated with the DiFX correlator in Socorro \citep{Del07, Del11}. 

The correlated data were reduced and processed in AIPS \citep{Gre03} as a first step to performing phase and amplitude calibration. Single-source FITS files containing calibrated data were produced and then loaded into Difmap for preliminary model-fitting and imaging. A cell size of 0.1 $\times$ 0.1 milliarcsecond was used and the image was produced with natural weighting. The outcome of the VLBI observations and data analysis revealed an unresolved radio source associated with GRB 100418a having a correlated flux density of 890 $\pm$ 50 $\mu$Jy, where the error is the 1$\sigma$ RMS noise.

\begin{figure*}
\includegraphics[angle=0,scale=.50]{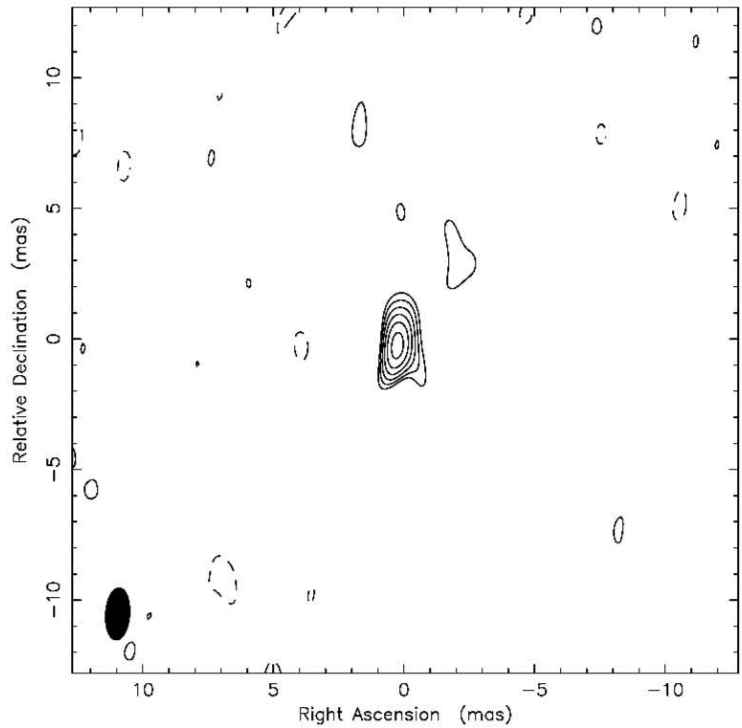}
\centering
\caption{GRB 100418a VLBI image. Observations at 8.4 GHz (2010 June 22). Contour levels: -10, 10, 20, 40, 80 \% of peak intensity; Peak flux density = 890 $\pm$ 50 $\mu$Jy (1$\sigma$); Beam size: 1.99 $\times$ 0.919 mas; Position angle -3.7$^\circ$.\label{fig2}}
\end{figure*}

Fig.~\ref{fig2} is the VLBI image of the radio afterglow of GRB 100418a at 8.4 GHz. It is interesting to note that the VLBA flux density matched almost exactly with the flux density obtained from the VLA observation on the same day, indicating that there was no missing flux due to the long baselines of the VLBA resolving out any structure in the afterglow.

The GRB 100418a radio afterglow position obtained from the VLBA observation is 17:05:27.092 $\pm$ 0.005, +11:27:42.24 $\pm$ 0.01 (Calibrator position: 17:06:20.4974, +12:08:59.794 with an uncertainty of 0.5 mas; http://www.astrogeo.org). The errors were estimated on the basis of the beamsize and signal-to-noise ratio (SNR) considerations. A systematic uncertainty of 0.06 mas was estimated in both coordinates based on the calibrator-target separation (which was 0.72$^o$ in this case) \citep{Pra06}. The VLBI position is consistent with the position of the optical counterpart detected by GROND with an error circle of 0.5 arcsec \citep{Klo10}, as well as the position of the radio counterpart detected by the ATCA and VLA.

The source was unresolved at the VLBA beam size of 1.99 $\times$ 0.919 mas. An upper limit on the angular extent of the source of $<$ 0.33 mas (1$\sigma$) was obtained by performing model-fitting in both $(u,v)$ and image plane using elliptical Gaussian models in AIPS. The upper limit ruled out that there was any larger scale structure at flux densities above the detection threshold at the time of VLBA observations of the afterglow at 8.4 GHz. Based on the size upper limit, the maximum physical size of $<$ 1.375 $\times$ 10$^{18}$ cm and the apparent expansion speed $<$ 8c were also determined. The VLBI observations of GRB 100418a were conducted about 65 days after the prompt emission and the source was found to be unresolved, so the derived upper limit on the apparent expansion speed was a long-term average assuming uniform expansion (mildly relativistic) and it does not conflict with possible highly relativistic expansion at early times.

The lower limit on the brightness temperature of the radio emission region was estimated as $T_b$ $>$ 2.95 $\times$ 10$^9$ K. The value was determined following \citet{Tay08}. This lower limit on the brightness temperature is high enough ($>$10$^7$ K) that the emission must be non-thermal. Therefore, the most likely emission mechanism is synchrotron emission resulting from the excitation of electrons during the interaction of the forward shock with the circumburst medium. Taking advantage of multi-frequency observations, the spectral index $\alpha$ (where $S_{\nu}$ $\propto$ $\nu^{\alpha}$) from the ATCA epoch closest to the VLBI observations was estimated as $\alpha$ = 0.5 $\pm$ 0.3, which indicates that the GRB radio afterglow was optically thick at 8.4 GHz. 

\section{Analysis and Interpretation}

GRB 100418a was ordinary in terms of the estimated energy release and the luminosity, but what made it an extraordinary GRB was the unusual afterglow emission behaviour i.e. the plateau phase in X-ray, re-brightening of the optical, and the late time rise and longevity of the radio afterglow. \citet{Zha06} proposed a model called the ``energy injection model'' to explain this kind of behaviour. This model was invoked by \citet{mar11} to answer questions concerning the mechanisms which could explain the unusual behaviour of the GRB 100418a afterglow.

The radius of the radiating GRB shell can be crudely approximated as R $\sim$ $\delta$tc, where $\delta$t is the time between the prompt emission and the peak of the radio afterglow and c is the speed of light \citep{Kat94, Kat97, Fra97}. From the VLA light curve, R is estimated to be $\sim$1.56 $\times$ 10$^{17}$ cm. This value obtained using this approximation is consistent with the predictions of the relativistic fireball model for GRB radio afterglows presented by \citet{Wax97a, Wax97b, Wax97c}. This model states that the radio emission associated with a GRB comes from a cone of the fireball along the observer's line-of-sight and R is therefore the apparent radius of the cone, which is suggested to be the emitting region. The brief decline and then the significant rise in the radio afterglow light curve (Fig.~\ref{fig1}) is indicative of the fact that there must be some sort of energy injection mechanism which refreshed the forward shock and in turn re-energized the fading afterglow.

\subsection{Theoretical framework}

Before attempting to explain the behaviour of GRB 100418a afterglow in the context of the energy injection model, it is important to put the observational signatures of the GRB 100418a afterglow in the context of the overall GRB phenomenon, and the standard blast wave model \citep{ree92, Mes93, Mes97a, Wax97a, Wij97}, which is still the most plausible physical framework, since the background physics proposed by this model has been repeatedly found to be in good agreement with GRB observations. The blast wave, or cosmological fireball model, proposes that a compact central engine produced as a result of the massive explosion of the progenitor star, launches a relativistic outflow releasing an enormous amount of energy (of the order 10$^{52}$ ergs). The outflow powers internal shocks \citep{ree94} having different Lorentz factors. Collisions between these internal shocks are thought to produce the GRB prompt emission. Following the internal shocks, a blast wave, also known as a forward shock or external shock, powered by the expanding ejecta, is driven into the circumburst medium, which is believed to produce the afterglow that is then observed at lower frequencies. The forward shock accelerates the electrons in the circumburst medium to relativistic speeds and they emit synchrotron radiation as they move in the surrounding magnetic field, which can then be detected and monitored for long-term afterglow studies \citep{Mes97, Kul98a, Sar98, Wak98, Fra00}.

Studies have indicated that the various GRB afterglow phases seen in the light curves represent the transitions which the decelerating ejecta and the expanding blast wave might be going through. In other words, the variations in the afterglow light curves can help infer the state of the emitting region. Fig. 1 of Zhang et al. (2006) shows the five possible components that the light curve of a GRB X-ray afterglow can have based on its decay index $\alpha$ ($F_{\nu}$ $\propto$ $t^{-\alpha}$). \citet{mar11} showed that the X-ray light curve of GRB 100418a exhibits three distinct phases: a) steep decline ($\alpha$ = 4.18); b) the shallow plateau phase ($\alpha$ = --0.23); c) relatively normal decay ($\alpha$ = 1.37). Phases ${a}$ and ${c}$ match well with segments I and III of Fig. 1 of \citet{Zha06}, respectively. However, the second phase of GRB 100418a X-ray afterglow features a very shallow rise instead of a shallow decline phase that matches with segment II of the \citet{Zha06} synthetic light curve. 

The rapid early decay of the GRB 100418a light curves is common amongst most of the GRBs and is attributed to the ``GRB tail-emission'', that is the steeply declining prompt emission that comes from the internal shocks immediately after the burst \citep{Der04, Lia06, Zha06}. In general, the transition from declining prompt emission to early X-ray emission manifests itself as the transition from a sharp decline to a shallow decline in the afterglow light curve \citep{Tag05}.

The second component in the optical and X-ray light curves is the unusual plateau phase with a very shallow rise. This is a rare feature in terms of the duration and behaviour and it appears to be a manifestation of some sort of a continuous post-burst activity. This behaviour has only been seen in very few GRB X-ray afterglows (e.g. GRB 060729; \citealt{Gru07}).

The energy injection model proposes that if the external (forward) shock that produces the GRB afterglow is continuously fed with energy after the initial supply, the forward shock keeps getting refreshed to produce a multi-wavelength afterglow for a longer period of time. According to the model, the onset of this continuous energy injection manifests itself as the very shallow plateau phase in the afterglow light curve, which is what is seen in the case of the afterglow of GRB 100418a.

The total isotropic kinetic energy in the case of GRB 100418a, which powered the plateau phase in the X-ray afterglow, was estimated by \citet{mar11} as $E_{k,iso,f}$ $\ge$ 10$^{53}$ ergs given $t_i$ $\sim$ 600s and $t_p$ $\sim$ 50-90 ks, where $t_i$ is the time when the plateau phase starts, and $t_p$ is the time when the afterglow reached its peak before the transition to a steep decay phase. This energy is 100 times the initial isotropic energy ($E_{k,iso,i}$ $\sim$10$^{51}$; \citealt{mar11}) and is thought to be continuously injected into the forward shock, which then produced a long-term, slowly varying afterglow \citep{Dai98, Zha01, Zha06}. The energy budget is thus consistent with the energy injection model and plausibly explains the behaviour exhibited by the afterglow of GRB 100418a.

The following mechanisms can give possible explanations for the energy injection model:

\begin{itemize}

\item Ejecta with a wide range of Lorentz factors, transporting energy to the forward shock \citep{ree98, Zha06}.

\item Continued activity of the central engine producing a Poynting flux dominated flow \citep{Dai98, Zha06}.

\end{itemize}

In the first scenario, multiple spherical shells are ejected with a range of Lorentz factors.  They reach the forward shock at different times, continuously refreshing the forward shock to give rise to a longer-lasting and slowly-varying afterglow \citep{ree98}. This model can be described as a relation between the fractional mass $M_{ej}$, ejected with Lorentz factors above a certain value of $\gamma$, and the range of Lorentz factors of the shells of material:

\begin{equation}
M_{ej} (>\gamma) \propto \gamma^{-s} 
\end{equation}

\noindent where \emph{s} is the mass outflow index and $\gamma$ is the Lorentz factor.

\subsection{Observational constraints}

The GRB 100418a VLBI observations led us to test this model by determining an upper limit on the long-term average of the Lorentz factor of the ejecta $\gamma$ $<$ 7, which we were able to obtain by estimating the upper limit on the apparent expansion speed as $<$ 8c. The upper limit on the expansion speed was determined using the upper limit on the source size obtained from the VLBA data, and estimating the value of angular diameter distance assuming a $\Lambda$ cosmology with $H_0$ = 71 km$s^{-1}$ Mp$c^{-1}$, $\Omega_M$ = 0.27 and $\Omega_{\Lambda}$ = 0.73 \citep{Rai01,Tay04}, at the redshift (z = 0.6235) of GRB 100418a. Using the formulation (Sedov-von Neumann-Taylor solutions) presented in \citet{Fra00}, the equivalent isotropic energy $E_0$ associated with the radio afterglow can be given as:

\begin{equation}
E_{0} \approx 4.4 \times 10^{50} \eta_1^{-2/17} \nu_*^{-0.56} a_*^{35/17} b_*^{15/17} d_*^{70/17}~ergs
\end{equation}

where $a_*$, $b_*$, $d_*$, $\eta_1$, $\nu_*$ are the parameters which define the afterglow model developed by \citet{Fra00} and are related to the fireball expanding at a subrelativistic velocity. Given the constraints (e.g. upper limits on physical size and Lorentz factor) obtained from the VLBI observations, the values of the parameters in Eq. 2 were calculated using the expressions given in the appendix of \citet{Fra00}. Those values were then used in Eq. 2 and the upper limit on the value of the afterglow (fireball) energy estimated to be  $E_0$ $\le$ 5.8 $\times$ 10$^{51}$ ergs. This value is two orders of magnitude less than the value of the isotropic kinetic energy estimated by \citet{mar11} because more than eight weeks after the prompt emission, the afterglow was in the radio regime and the accelerated particles had lost some energy, becoming less and less relativistic as the fireball was expanding in the circumburst medium.

The ejected mass $M_{ej}$ corresponding to the radio afterglow energy release can be estimated using the formulation presented in \citet{Pan02}, based on Einstein's equation relating relativistic mass and energy:

\begin{equation}
M_{ej} \approx \frac{E_{0}}{c^2\gamma_{0}}~kg
\end{equation}

The upper limit on the  value of mass corresponding to this energy release was determined as $M_{ej}$ $\le$ 0.46 $\times$ 10$^{-3}$$M_{\odot}$, where $M_{\odot}$ is the solar mass. 

\subsection{Test of the energy injection model}

\begin{figure*}
\includegraphics[angle=0,scale=.50]{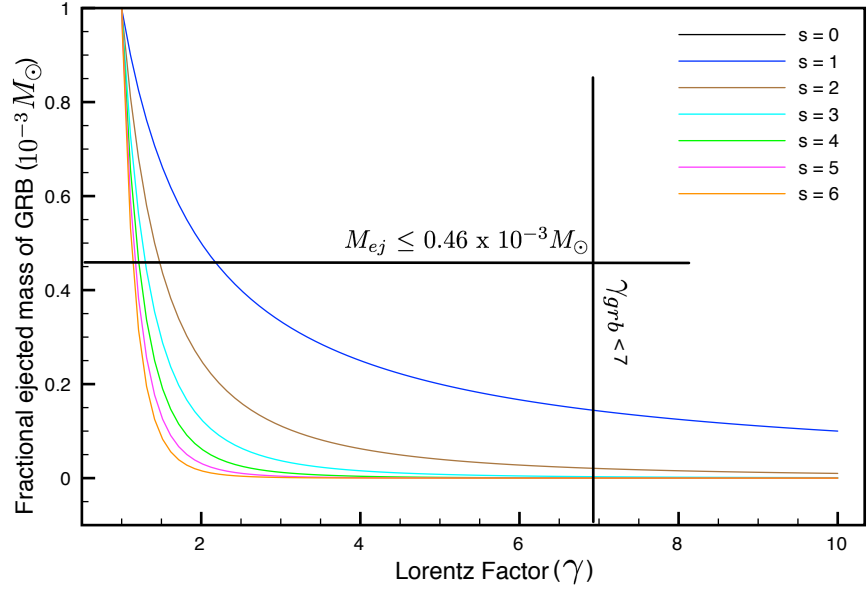}
\centering
\caption{The relationship between the ejected GRB mass $M_{ej}$ and Lorentz factor $\gamma$ (Eq. 1) plotted for different values of mass outflow index ${s}$ (Rees \& Meszaros 1998; Zhang et al. 2006). The estimation of upper limits on $M_{ej}$ and $\gamma$ indicates that it is only within the window shown in the figure that the range-of-Lorentz-factor hypothesis for energy injection may be true. That is, only for slow moving ejecta with lower values of $\gamma$. \label{fig3}}
\end{figure*}

The constraints on $\gamma$ and $M_{ej}$ obtained from the VLBA observations allowed us to test the model relating the ejected mass and Lorentz factor (Eq. 1) against the observations from which it can be concluded that in case of GRB 100418a, the range-of-Lorentz-factor postulate of the energy injection model is in fact only valid for a small range of Lorentz factors i.e. $\gamma$ $<$ 7, and that 65 days after the GRB 100418a prompt emission, the fractional mass outflow and the energy release were indeed dominated by slow moving ejecta having values of $\gamma$ $<$ 7.That is, it is only the slow moving shells that contributed to the longevity of the afterglow, and there must have been another mechanism, such as extended activity of the central engine, which was driving the continued supply of energy to the forward shock in the beginning and in turn powering the long-lasting afterglow. The VLBA Lorentz factor limit ($\gamma$ $<$ 7) is model-independent since it is a direct observation. Fig.~\ref{fig3} shows the plots of Eq. 1 for a range of mass outflow indices 0 $<$ s $\le$ 6. The mass outflow is dominated by slow moving shells for s $>$ 1 \citep{ree98} and for s $>$ 6, the curves start to converge indistinguishably. It, therefore, is only a small window provided by the limits on $\gamma$ and $M_{ej}$ within which the observations are consistent with the wide range of Lorentz factor model \citep{ree98, Zha06} and the model does not hold for higher values of $\gamma$. Our conclusion thus strengthens the argument in favour of the continued activity of the central engine as the process that primarily powered the prolonged afterglow.

In the case of continued activity of the central engine, the existing theoretical framework suggests that there could be at least some long-duration GRBs that are produced when a massive progenitor collapses into a neutron star or a highly magnetized pulsar (the proto-magnetar model; e.g. \citealt{Met11}) . During the birth of a pulsar or magnetar, the relativistic outflow produces internal shocks closer to the central engine which could produce the prompt emission, with the energy transported to a distance feeding the external shocks which produce the GRB afterglow. The central engine may then start to spin down and lose its energy by magnetic dipole radiation. In the presence of a strong electro-magnetic field, the energy is transported via a relativistic magnetohydrodynamic wind as Poynting flux energy which keeps refreshing the external (forward) shock, due to which it decelerates less slowly in the circumburst medium, resulting in continued emission of afterglow radiation \citep{Dun92, Tho94, Zha01, Zha06}. 

Another possible scenario describes the continuous late time accretion of ``fall-back'' material onto a black hole as the primary mechanism powering energy injection. It has been proposed that late-time hyper-accretion is sustained by distant chunks of material which eventually end up in the vicinity of the accreting black hole. This material is accreted onto the black hole refreshing the accretion and fueling the long-term energy injection (\citealt{Per06, Kum08a, Gen13} and many references therein).

Discriminating between these two models to explain the continued activity of the central engine is beyond the scope of this work. However, the pulsar model could be further explored by a comparison between the pulsar rotational energy and the GRB energy and the similarities between the pulsar environment and the circumburst medium (e.g. \citealt{Dai98}), with the aid of regular observational campaigns. The black hole hyper-accretion model can be further investigated by searching for observational signatures of an accretion disc (e.g. \citealt{Per06, Can11}).

In the case of GRB 100418a, the continued injection of energy into the forward shock can also explain the long-term radio emission. The forward shock may have been continuously fed with energy from the activity of the central engine, heating up the gas in the surrounding medium and accelerating electrons to relativistic velocities in an optically thick region. Therefore, the afterglow kept rising until it reached its peak and lasted much longer than many other GRBs. With the expansion and deceleration of the emitting sphere, the radio afterglow eventually started fading slowly.  In addition to the observed results, a simple mathematical formulation of the standard blastwave model for GRB afterglows \citep{Gra03, Tay04, Pih07} was used to predict the theoretical upper limit for angular size of GRB 100418a, based on a standard cosmology and on the day it was observed with the VLBA, $<$0.04 mas. This upper limit is much smaller than the upper limit estimated using the data from the observations ($<$0.33 mas) but is consistent with the conclusion that the afterglow was unresolved even more than 60 days after the prompt emission, indicating that the forward shock or the shells of material weren't expanding fast enough (i.e. slow moving ejecta) that it could be resolved. 

Since both observational and theoretical estimation of the limits on the size and speed of the afterglow lead us to the conclusion that the range of Lorentz factors model is only partially valid, it is likely that the long-term activity of the central engine is required to explain the longevity of the afterglow.

The slow rate of expansion of the afterglow implies that the material the ejecta were interacting with was dense, consistent with the persistent and strong radio emission. But due to the faintness of the GRB 100418a host galaxy (e.g. from the SDSS images; \citealt{Mal10}), it is not possible to determine whether the GRB occurred in a high density region (close to the centre) of the galaxy or not.

\section{Conclusions}

GRB 100418a was an unusual GRB in terms of the duration and properties of its afterglow. The afterglow light curves showed a rare ``plateau phase'' which is thought to be a signature of the physical process associated with the longevity of the afterglow. A long-term observational campaign was carried out with the VLA, ATCA and VLBA to monitor the afterglow in the radio band. The behavior of the ATCA + VLA light curve shows that the radio afterglow started to rise after a brief period of decline indicating some kind of energy injection re-energized the forward shock which in turn produced a long-lasting afterglow. 

We used our radio observations to test one of the postulates of the energy injection model, deriving constraints on the outflow Lorentz factor and the GRB afterglow mass.  It was found that the upper limits only partially support the notion that a range of Lorentz factors might have continuously supplied energy to the forward shock to produce long-lasting afterglow and that it could only be true for slow moving shells having Lorentz factors $\gamma$ $<$ 7. Therefore, in the case of GRB 100418a, it is inferred that it was only the slow-moving shells that could have contributed to the continuous energy injection to the forward shock at late times. There must have been significant contribution to the long-lasting afterglow from some other mechanism e.g. the continued activity of the central engine due to which the forward shock continued to get sufficient energy to accelerate particles from the beginning and the slow moving shells caught up with it at late times. Our GRB 100418a monitoring suggests that it is very important to keep searching for and not to miss a GRB afterglow as unusual and long-lasting as GRB 100418a, which should then be continuously monitored with frequent sampling in order to further test GRB theories.

\acknowledgments

This work made use of ATNF CASS' ATCA, and NRAO's VLA and VLBA. We thank the schedulers and all of the support staff at NRAO and ATNF CASS. The Australia Telescope Compact Array is part of the Australia Telescope National Facility which is funded by the Commonwealth of Australia for operation as a National Facility managed by CSIRO. The National Radio Astronomy Observatory is a facility of the National Science Foundation operated under cooperative agreement by Associated Universities, Inc. The research was supported in part by the National Natural Science Foundation of China (NSFC) under No.11073042, National Basic Research Program of China (973 Project 2009CB824800). ZW is a Research Fellow of the One-Hundred-Talents project of Chinese Academy of Sciences.

\clearpage

\begin{table*}
\centering
\begin{minipage}{140mm}
\caption{VLA, ATCA and VLBA observations of GRB 100418a. The GRB was detected on all epochs ($>$3$\sigma$; except the 4.9 GHz session on 15/09/2011 and the 7.9 GHz session on 21/08/2012 when the detections were $<$3$\sigma$). $\delta$T is the number of the days since prompt emission.}
\begin{tabular}{@{}llrrrrlrlrlrrrrrrrrrrrrrrrrrrrrrrrrrr@{}}
\hline
Date     & Instrument  & $\delta$T   & Frequency & Configuration& Flux density &  RMS \\
              &            & (days)   &  (GHz) &   & ($\mu$Jy\,beam$^{-1}$)  & ($\mu$Jy\,beam$^{-1}$)\\ \hline 
 20/04/2010 & ATCA & 2 & 5.5 &   6A & 865 & 120   \\
 20/04/2010 & ATCA & 2 & 9.0 &   6A & 1273 & 96   \\
 21/04/2010 & VLA & 3 & 8.46 &  D  &  458 & 20  \\
 24/04/2010 & VLA & 6 & 8.46 &   D & 289 & 22  \\ 
 30/04/2010 & VLA & 12 & 8.46 &   D & 516 & 22  \\ 
 06/05/2010 & VLA & 18 & 8.46 &    D & 537 & 20    \\
 14/05/2010 & VLA & 26 & 8.46 &   D & 847 & 23    \\
 21/05/2010 & VLA & 33 & 8.46 &   D & 1028 & 37    \\
 26/05/2010 & ATCA & 38 & 5.5 & 6C & 900 & 80  \\
 26/05/2010 & ATCA & 38 & 9.0 &   6C & 1370 & 180   \\
 05/06/2010 & VLA & 48 & 8.46 & D& 1454 & 21  \\
 15/06/2010 & VLA & 56 & 4.95 &   D & 440 & 76 \\ 
 15/06/2010 & VLA & 56 & 8.46 &   D &  923 & 32  \\
 22/06/2010 & VLBA & 65 & 8.4 &  & 890 & 50 \\ 
 26/06/2010 & ATCA & 67 & 5.5 &   6C & 1270 & 120   \\
 26/06/2010 & ATCA & 67 & 9.0 & 6C & 1600 & 200  \\
 29/06/2010 & VLA & 72 & 4.95 &    D& 513 & 44  \\ 
 07/07/2010 & VLA & 80 & 8.46 &   D & 526 & 41 \\ 
 22/07/2010 & VLA & 95 & 8.46 &  D   & 641 & 37    \\
 22/07/2010 & VLA & 95 & 4.95 &  D  & 404 & 50    \\
 22/09/2010 & VLA & 157 & 8.46 & DnC  & 366 & 53    \\
 22/09/2010 & VLA & 157 & 4.95 & DnC & 196 & 64  \\
 06/04/2011 & VLA & 353 & 4.9 &  B  & 311 & 25   \\
 06/04/2011 & VLA & 353 & 7.9 &  B & 332 & 20  \\
 15/09/2011 & VLA & 391 & 4.9 &  A-D  & 205 & 82   \\
15/09/2011 & VLA & 391 & 7.9 &  A-D & 105 & 26  \\
21/08/2012 & VLA & 721 & 4.9 &  B  & 148 & 28   \\
21/08/2012 & VLA & 721 & 7.9 &  B & 81 & 34  \\
 
 \hline
\label{obs}
\end{tabular}
\end{minipage}
\end{table*}

\end{document}